\newcommand*{\LONG}{}%
\pdfoutput=1

\ifdefined\LONG
\documentclass[11pt]{article}
\usepackage{amsfonts,amsmath,amssymb,latexsym,comment,amsthm}

\usepackage[letterpaper]{geometry}

\usepackage{mathrsfs}
\usepackage{times}

\else 
\documentclass{sig-alternate-2013}
\usepackage{amsfonts,amsmath,amssymb,latexsym,comment }

\fi

\usepackage{qtree}
\usepackage{epsfig}
\usepackage{graphicx}
\usepackage{subfigure}

\usepackage{xspace}
\usepackage{cases}
\ifx\pdftexversion\undefined
\usepackage[colorlinks,linkcolor=black,filecolor=black,citecolor=black,urlcolor=black,pdfstartview=FitH]{hyperref}
\else
 \usepackage[colorlinks,linkcolor=blue,filecolor=blue,citecolor=blue,urlcolor=blue,pdfstartview=FitH]{hyperref}
\fi
\usepackage[ruled,boxed,commentsnumbered,noresetcount]{algorithm2e}
\usepackage{enumitem}

\usepackage{pifont}

\def\squarebox#1{\hbox to #1{\hfill\vbox to #1{\vfill}}}

\newtheorem{theorem}{Theorem}

\newtheorem{lemma}{Lemma}

\newtheorem{corollary}{Corollary}

\newcommand{\namedref}[2]{\hyperref[#2]{#1~\ref*{#2}}}

\newcommand{\algref}[1]{\namedref{Algorithm}{#1}}

\newcommand{\lemmaref}[1]{\namedref{Lemma}{#1}}

\newcommand{\lref}[1]{\namedref{Line}{#1}}

\newcommand{\MOBfm}{\mbox{MOB\!$f\!m$}\xspace}
\newcommand{\MOBtt}{\mbox{MOB$tt$}\xspace}
\newcommand{\MAOBt}{\mbox{MAOB$t$}\xspace}

\newcommand{\cosend}{\mbox{\sc co\_send}\xspace}

\newcommand{\ie}{\emph{i.e.,\xspace}}

\newcommand{\tri}[1]{\left<#1 \right>}

\def\beginsmall#1{\vspace{-\parskip}\begin{#1}\itemsep-\parskip}
\def\endsmall#1{\end{#1}\vspace{-\parskip}}

\newcommand{\F}{\mathcal{F}}

\newcounter{todocounter}

\newcommand{\tb}{\makebox[0.6cm]{}}
\newcommand{\due}{\makebox[1cm]{}}

\newcommand{\hide}[1]{}

\newcommand{\commentout}[1]{}

\newcommand{\M}{\mathcal{M}}

\newcommand{\I}{\mathcal{I}}

\newcommand{\ccore}{{\small\textsc{common\!\_core}}\xspace}

\newcommand{\simcoin}{{\small\textsc{sim\!\_coin}}\xspace}

\begin{document}

\title{Some Garbage In - Some Garbage Out: Asynchronous $t$-Byzantine as Asynchronous Benign $t$-resilient system
with fixed $t$-Trojan-Horse Inputs}
\author{Danny Dolev (HUJI)\footnote{Email: danny.dolev@mail.huji.ac.il.} \ and
Eli Gafni (UCLA)\footnote{Email: gafnieli@gmail.com.}\\[1.2ex]
}

\date{}

\maketitle

\thispagestyle{empty}

\setcounter{page}{0}

\section*{Abstract}

We show that asynchronous $t$ faults Byzantine system is equivalent to asynchronous $t$-resilient system,
where unbeknownst to all,  the private inputs of at most $t$ processors were altered and installed by a malicious oracle.

The immediate ramification is that dealing with asynchronous Byzantine systems does not call for new topological methods,
as was recently employed by various researchers: Asynchronous Byzantine is a standard asynchronous system with an input caveat.
It also shows that two recent independent investigations of vector $\epsilon$-agreement in the Byzantine model, and then in the fail-stop
model, one was superfluous - in these problems the change of $t$ inputs allowed in the Byzantine has no effect compared to the fail-stop case.

This result was motivated by the aim of casting any asynchronous system as a synchronous system where all processors are correct and
it is the communication substrate in the form of message-adversary that misbehaves. Thus, in addition, we get such a characterization 
for the asynchronous Byzantine system.

\newpage

\section{Introduction}

Recently, \cite{hybrid-arxiv}, extending \cite{AG15}, we have shown that for output colorless tasks - tasks whose correctness is preserved if in a correct output the output of $p_i$ is replaced by the output of $p_j$, e.g., $\epsilon$-agreement - an asynchronous $t$-resilient model is equivalent to synchronous network with $t$-mobile message adversary.  That is,
an adversary that each round can choose \emph{any}
$t$ nodes and remove some of the messages they send. 
This motivated us to do the same for asynchronous Byzantine system. 
It turns out that we need to tolerate the possibility that $t$ input values are faked.  

How did the  $t$ resilience motivate us to examine $t$ fault asynchronous Byzantine?
In a $t$ resilient system we can solve $n-t+1$-set consensus. A recent result of Generalized Universality \cite{Rachid} implies that consequently one can view execution as a free for all and  guarantee that at least $n-t$ threads will advance from round to round. The only association between processors and threads is in activating a thread by posting an input to the thread and departing, once an output of the thread has been determined. In between, the advancement of threads is a cooperative process among the processors with no special privileges to a ``thread ownership.'' Byzantine faults are static faults, we have fixed $t$ processors which can exhibit such faults. In light of the disassociation of processors and threads in the $t$-resilient case, can't we thwart the effect of Byzantine processors by making every step of advancing the threads a cooperative effort?

The main result of the current paper is that indeed this is the case. A Byzantine behavior within the execution can be completely neutralized by making
the execution of the threads a cooperative effort, rather than associating between processors and threads while executing. Thus, effectively
we make the notion of 
\emph{malicious} processors a misnomer.
A gist of such an idea did occur to Coan almost 3 decades ago \cite{coan88}. But as we elaborate later, Coan tried
to precisely emulate the messages of a given protocol, and thus the main equivalence as models got lost.

Although motivated by Generalized Universality we give a tailor made argumentation to the fact that the execution of asynchronous Byzantine is effectively an execution of a standard asynchronous system,  with a caveat. We have a Byzantine adversary that can tamper with at most $t$ inputs. To model this we envision that in the first round (``Round 0'') a valid set of inputs is sent to processors. 
 An adversary can intercept $t$ of these inputs and change them.
From then on (``Round 1'' on), we model the asynchronous $t$-resilient system as a synchronous one with a $t$-mobile message adversary.
In each round we do not have ``incorrect'' processors, who are ``malicious,'' etc., all we have are some $t$ processors that might
be experiencing a send malfunction in that round. Moreover, this model enables us to talk about all processors achieving the objectives of the protocol since the failure surfaces only in sending out some messages.

This result brings the asynchronous Byzantine model into the mainstream as we can analyze the system as a normal $t$-resilient system. Albeit, we do have to contend with inputs that might be inconsistent and ascribe some meaning if possible to the outputs. 
Two recent results are striking in this regard. In \cite{Mendes2015} the authors have formulated an ingenious task whose inputs are points $R^d$ and asked as a function of $t$ and $n$ whether the problem of $\epsilon$-agreement that is in
the convex-hull of every $n-t$ inputs is solvable in the asynchronous Byzantine model. 
At the instigation of the second author who felt that the problem has nothing to do with the Byzantine setting, 
in \cite{TV14} the same problem was analyzed in the benign case by some of the original authors in \cite{Mendes2015} and lo and behold they obtained the same results. 
Our paper here shows this was not a fluke.

In fact, to give more credence to our contention that we are not the only ones that
should be surprised by the correspondence of the Byzantine $t$ faults and the $t$-resilient systems, in \cite{Mendes2014},
the authors treat Byzantine asynchronous systems as a new animal and develop special topological methods for it.
Our current paper stops such a research direction in its track.

The main challenge in obtaining the result is to show that any deterministic protocol that solves a problem in the $t$-mobile message adversary synchronous system can be simulated by a protocol running in the traditional $t$ resilient asynchronous Byzantine  system.  
Naturally, the adversary seems to have much more liberty in the traditional $t$ resilient Byzantine model, since it can delay $t$ correct processors, and arbitrarily control the $t$ Byzantine nodes.
Traditionally, researchers developed methods to exchange the set of values being received from each processor in order to ensure consistency at every stage of the protocol. Such methods achieve consistency, but still enable a faulty node to claim to receive values that were not sent, and extra methods are required to test consistency over the history of the protocol
(as in our discussion of \cite{coan88} below). 

Our approach is to eliminate sending values beyond the initial inputs. Instead, every round each processor sends to others the list of processors it heard from in the previous round.  Sending only this list drastically reduces the ability of the adversary to influence the state of the rest of the processors.  With that in mind, the protocol is simulated locally at each processor.  Each processor can determine which values each other processor needs to send and receive in each round, given the set of processors it claims to hear from. 
Schneider~\cite{FTState90} studied the idea of using a state machine approach to implementing 
fault-tolerant services.  His approach is to instruct all replicas to run the same state machine, and to agree on the inputs to the replicas.  We take the idea further and instruct each processor to simulate the protocols of all other processors, and our simulation protocol ensures that all replicas at all correct processors apply the same sequence of steps.  This resembles, to some extent,  the approach of Gafni and  Guerraoui~\cite{Rachid}.

There were several attempts to simplify the $t$ resilient asynchronous Byzantine model  through ideas of simulating simpler models.  Attiya and Welch~\cite{DistComp} reduced the problem to Identical Byzantine.
The pioneering work of Bracha~\cite{Bracha:1984,Bracha:1987} was focused on improving the probabilistic protocol of Ben-Or~\cite{BenOr83} from $n/5$ to $n/3$ and in order to do so Bracha developed a basic tool to limit the power of the Byzantine adversary,  The simulation we introduce in the paper makes use of this tool as part of the building block we introduce.
Srikanth and Toueg~\cite{Srikanth87} considered simulating the power of a signature scheme to limit the Byzantine adversary, both in a synchronous system and an asynchronous one. Neiger and Toueg~\cite{neiger88} introduced direct simulations between models in order to solve consensus, but their simulations are limited to synchronous models.

Coan~\cite{coan88} technique comes the closest to our result.
Coan was interested in taking a given algorithm written for the asynchronous fail-stop model with $t<1/3$ and running
it in an environment of $t$ Byzantine processors. This is a more ambitious goal that what we present here.
At first, cut the ramification of our paper is ``for tasks that are \emph{immune} to a change of at most $t$ inputs,
whatever is solvable asynchronously with $t$-benign faults is solvable with $t$ Byzantine faults.''

Coan pays for the ambition. The algorithm for the fail-stop environment has to be written in a specific form that is
not shown to be universal and encompass every protocol, and then run through a compiler to validate ``message correctness,''
and ``filter out'' incorrect messages. Thus, in hindsight, we believe Coan ideas can be tweaked to get our result,
but this is in hindsight. 
The results were surprising to us as it should be to most researchers evidenced by the 
recent duplicate works mentioned above.

Last but not least, we do not address the question for the randomized environment when processors flip coins 
(remember, our processors are correct, only that the communication subsystem interfere with them). Coan faces the
question then how to define ``correct message.'' We do not face this problem since ours is full information and
about the ``communication pattern.'' In that case, we need all processors to agree on individual processor's claimed coin
output, and we still face the problem that decided coin values might be biased, unlike the fail-stop case.
Nevertheless as we show in the appendix if we go down to $t <1/4~ n$ we can deal with randomized
algorithms too. The case of $t <1/3~ n$ is an open question.

Last, Coan do address the falsified input question and assumes some ``correctness predicate.''
We leave it to the protocol designer to address the question what to do with a combination of inputs that is
not a valid input combination.

Several papers discuss methods to simulate shared memory in  a message passing system with Byzantine processors. These papers, as a by-product, limit the power of the Byzantine adversary.
Malkhi and Rieter~\cite{Malkhi98DC} use a simulation assuming that information from correct nodes is self-verifying (e.g., digitally signed). They also  defined opaque masking quorum systems, which allows simulating a shared register without assuming that data values are self-verifying, but assumed a higher ratio of non-faulty nodes.
Abraham, Chockler, Keidar, and Malkhi~\cite{DiskPaxos2005} present a simulation that  provides weaker shared-memory properties and  terminates only if the number of writes is finite.
Aiyer, Alvisi and Bazzi~\cite{Aiyer2007} use a secret sharing scheme to simulate an atomic register, with Byzantine readers and (up to one-third of) Byzantine servers.

\section{Problem Statement and Models}
We assume a set of $n$ processors 
$\Pi=\{p_1, p_2, ..., p_n\}$. 
The paper focuses on the equivalence between a $t$ resilient asynchronous Byzantine message passing system and a synchronous message passing, \MOBfm, system.
The \MOBfm, called \emph{mobile omission Byzantine $n,f,m$}, is a synchronous point-to-point message passing system, with
an adversary that can replace the input values of a set $S^f\subset \Pi$  of $f$ processors.
In addition in each synchronous round, the adversary can choose a set $S^m$ of $m$ processors and remove some 
or all of the messages sent by processors in $S^m$. 


We  prove that regarding deterministic protocols, the  \MOBtt system, \ie\  $m=f=t$, is equivalent to the classical asynchronous message passing model with $t$ Byzantine faults, for $n>3t$.
Obviously, each run in a \MOBtt system is a possible run in an asynchronous Byzantine system. Therefore, what we need is to prove that any deterministic protocol running in the \MOBtt model can be simulated by a deterministic protocol in  an asynchronous Byzantine system. 

We assume that each processor has an input value.  
A deterministic message passing protocol $P$ that solves a problem  in the \MOBtt system runs for a given number of rounds of message exchange and by the end of the protocol run each processor produces an output.  
For simplicity assume that in the first round of running protocol $P$ processors are expected to share their input values.
\footnote{Protocols in which the inputs are secrets can't be simulated using the technique we present.}  
For convenience, we assume that all messages sent in a given synchronous round are tagged by a counter indicating the round number. 
Any protocol $P$ can be viewed as a transition function $\F(M_{r-1},S_p,r)$ that instructs each processor $p$ in a round $r$, given  its current state  $S_p$, and given the set of messages, $M$, received in the previous round, which actions to take in the current round.  
An action is what message to send to which processor and whether to produce an output.  $M_0$ of the first round is just the input value.
Thus, the state of the protocol at each round is a function of the initial input and the sequence of sets of messages received in all previous rounds.

Observe that any processor $q$ that receives the  sequence of sets of messages  $M_0, ... M_{r-1}$ that were received by $p$ in previous rounds can determine what message $p$ should send it in  round $r$. 
Moreover, $q$ can also know what message any other processor should receive from $p$ in round $r$.
We take advantage of these observations in the simulation.

The challenge  is to turn the Byzantine processors into processors that behave consistently with the protocol.
The breakthrough idea is that instead of asking processors to send  values we instruct them to send only the set of processors  they received messages from in the previous round. Each processor uses this information to locally simulate the state of each other processor and to determine what messages each processor should have received and should have sent in each round.

As a step toward the result we first prove that $t$-resilient asynchronous Byzantine system is equivalent to a synchronous system that is similar to \MOBtt, called \MAOBt, in which the adversary replaces the input values of a set of size $t$ and from that point on in each round it can drop messages sent by any processor, as long as it does not drop more than $t$ incoming messages to any processor.

\section{Simulating a protocol in a MAOBt system}

In the simulation, we make use of several building blocks.  The idea behind the simulation is to completely simulate the protocol at each processor. The first technique employed is to make sure that everyone commits to the message it sends in each round in a way that if any processor accepts a message $m$, everyone will eventually accept  all $m$'s  causally ordered prior messages followed by the same message $m$. 
The second technique is to locally simulate the state of the protocol at every other processor, so we know what values should be sent and which should be received.
%

The first building block, \cosend, is the {\em Causally Ordered Reliable Broadcast primitive} (inspired by~\cite{Bracha:1984,DistComp}). 
The second primitive resembles~\cite{Rachid}, each processor is running locally the protocol's state machine of each other processor, according to the messages being received, to determine what the protocol instructs each processor to do.

Let $P$ be a deterministic message passing protocol that is executed  in a \MAOBt system.  We will show a simulation of it in a $t$ resilient asynchronous Byzantine system.  In the simulation, in the first round each processor, $p$, uses $\cosend(1,p)$ to broadcast  its input value, $\I$.  
In each subsequent round, $r$, each processor, $p$, uses $\cosend(r,p)$  to broadcast  to everyone the set of processors, $\pi_{r-1}$, from which it received messages in the previous round. 

We start with an overview of the simulation protocol.
Each processor maintains locally $n$ protocol state machines, $SM_i$, $1\le i\le n$.
When it accepts via \cosend a $\tri{1,\I}$ from a processor $p_i$, it initiates state machine  $SM_i$ with the input value $\I$.

When it accepts via \cosend a $\tri{2,\pi_{1}}$ from a processor $p_i$, it uses the initial values of all $q_j\in\pi_1$  as the set of  input values to $SM_i$, to determine what values processor $p_i$ should send to every other processor in round $2$. 
Since \cosend implements Causally Ordered Reliable Broadcast, before the processor processes $\tri{2,\pi_{1}}$  from a processor $p_i$, it already accepted and processed all $\tri{1,\I}$ messages from all processors in $\pi_{1}$.

Now recursively, when it accepts via \cosend a $\tri{r,\pi_{r-1}}$ from a processor $p_i$, it uses the  values every $q_j\in\pi_{r-1}$ should have sent in $r-1$ to $p_i$ according to $q_j$'s state machine $SM_j$ at round $r-1$  as values received by $SM_i$ in the previous round (round $r-1$)  to determine what values processor $p_i$ should send to every other processor in round $r$. 

The simulation protocol, presented in \algref{figure:state-sim}, maintains  three data structures.  
The set $\M$ contains the messages that were received via \cosend and that are not processed yet, there is at most one
such message per round per sender.  
Each entry in $\M$ contains a round number, say $r$, a processor ID, say $p_i$, and the set of processors' IDs, from which processor $p_i$ claims to have received messages in round $r-1$.

The set $\bar\M$ contains the list of processes whose messages were already processed. 
Each entry in $\bar\M$ contains a round number, say $r$, a processor ID, say $p_i$, indicating that round $r$ message from $p_i$ was received and processed. Every processor that processes a round $r$ message from $p_i$ processes the identical message.

The \cosend properties imply that when an entry is added to $\M$, all casually prior entries were already accepted and processed, and as such are reflected in the respective state machines (as we explain later). 
Therefore, each message in $\M$ can be processed independently, since there are no causal dependencies among them.
Processing a message is just applying it to the state machine of the sending processor, using the current state of the state machines of all
the processors it claimed to receive their messages in the previous round.
Once a message is processed it is removed from $\M$ and added the $\bar\M.$
Observe that the simulation may indicate that at a certain round some processor is not sending a value to some other processor, then in such a case no such value is produced as an input to the relevant state machine.

The third data structure ($accept$) is the set of processors whose messages were accepted in the given round. 
Let $\pi_r$ be the set of all the processors whose messages were accepted by $p$  via   \cosend during round $r$.   Once $|\pi_r|\ge n-t$, processor $p$ uses $\cosend(r+1,p)$ to broadcast $\pi_{r}$. After broadcasting this message processor $p$ continues to accept previous rounds' messages via  \cosend and continues to apply them to the various state machines.
Each processor continues this process, outputs its output, and sends messages until its state machine halts.\footnote{A the processor continues to participate in the \cosend protocols of other processors even after it halts.}  

\begin{algorithm}[!ht]
\footnotesize
\SetNlSty{textbf}{}{.}
\setcounter{AlgoLine}{0}
\lnl{line:def}  \     {\bf set}  $\forall k$ $accept[k]:=\emptyset$; \hspace{2mm} \hfill\textit{/*  the sets of accepted senders at various rounds; executed at processor $p$ */}\\
\lnl{line:def2}  \     {\bf set}  $\bar\M:=\emptyset$;\hspace{2mm} \hfill\textit{/*  the set of processed messages */}\\
\lnl{line:def3}  \     {\bf set}  $\M:=\emptyset;$\hspace{2mm} \hfill\textit{/*  the set of accepted  messages that were not processed yet  */}\\

\ \\
\lnl{line:step1} \    {\bf invoke }  $\cosend(1,p)$ to broadcast $\I$;\mbox{\ }\hfill\textit{/* broadcast the input value, a processor sends  also to itself */}\\
\lnl{line:init-r} \    $r:=1;$\hfill\textit{/* the round number */}\\
\ \\
\nl \    {\bf do }  {\bf until} $SM_p$ halts:\\
\lnl{line:collect}\    \tb {\bf wait until } $|accept[r]|\ge n-t$ and $p\in accept[r]$; \hfill\textit{/* participate in all $\cosend(\ell,*),$ $\ell \le r$, protocols */}\\
\lnl{line:core}\ \\
\lnl{line:send} \    \tb {\bf invoke } $\cosend(r+1,p)$ to broadcast $accept[r]$;\mbox{\ }\hfill\textit{/* broadcast the accepted set in round $r$ */}\\
\lnl{line:inc}\    \tb $r:=r+1$;\\
\lnl{line:end} \    {\bf end.} \\
\ \\
\lnl{line:r1}  {\underline {\bf In the Background:}} Execute for each $\tri{r',p_i,\pi}\in\M$: \hfill\textit{/*   message received from $p_i$ for round $r$   */} \\
\lnl{line:init-rec}    \tb {\bf if} $r'=1$ {\bf then start } $SM_i$ with input $\pi$;\hfill\textit{/* start a SM with the initial input */}\\
\lnl{line:init-rec2}     \tb {\bf if} $r'>1$ {\bf then } \\
\lnl{line:rec-m}  \tb\tb {\bf let} $M:=\{m_j\mid p_j\in\pi \mbox{ and } SM_j[r'-1] \mbox{ sends } m_j \mbox{ to } p_i\}$; \hfill\textit{/* the messages $p_i$ should have received */}\\
\lnl{line:apply-m}  \tb\tb$SM_i[k]:=\F(M,SM_i[r'-1],r')$; \hfill\textit{/* apply  protocol $\F$ to determine the next state of $SM_i$ */}\\
\lnl{line:r3}    \tb $\M:=\M\setminus \tri{r',p_i,\pi}$;\\
\lnl{line:r4}     \tb $\bar\M:=\bar\M\cup \tri{r',p_i}$;\\
\lnl{line:add}       \tb   $accept[r']:=accept[r']\cup\{p_i\}$. \\
\caption[caption]{Simulating a deterministic protocol of a \MAOBt system  \\\hspace{\textwidth}\mbox{\ \hspace{0.83in}} in an asynchronous Byzantine system with $n>3t$}\label{figure:state-sim}
\end{algorithm}

\begin{algorithm}[!t]
\footnotesize
\SetNlSty{textbf}{}{.}
\setcounter{AlgoLine}{0}
\ \hfill\textit{/* executed by processor $p$ with  sender  $s$ in round $r$, invoked once per round */}\\
 $\M$ and $\bar \M$ are globally maintained sets\\
 \ \\
 \nl    {\bf let}  $V$ be the set of $m_1$ and $m_2$ protocol messages received;\hspace{2mm}\ \hfill\textit{/* each processor sends also messages to itself */}\\
\ \\
\lnl{line:step0}    {\bf Init:}  {\bf if}  $p=s $ {\bf then}  \\
\lnl{line:step0a} \tb send $v_s$ to all;\mbox{\ }\hfill\textit{/* $s$ is the sender and $v_s$ the value it broadcasts */}\\
\lnl{line:step0b}  \tb  {\bf Accept:} $\M:=\M\cup\tri{r,s,v_s}$.\hfill\textit{/*   accept message $v$ from self  sent in round $r$  */} \\
\ \\
\lnl{line:co-exec1}    {\bf Upon receiving a protocol message:} \\
\lnl{line:co-exec2}    \tb {\bf case}  received $v$ from $s$:  \\
\due send $m_1(v)$ to all;\hfill\textit{/* executed at most once per protocol invocation */}\\
\ \\
\lnl{line:co-exec3}   \tb {\bf case}  $V$ contains $m_1(v)$ from  $n-t$ different processors or $m_2(v)$ from  $t+1$ different processors: \\
\due    send $m_2(v)$ to all;\\
\ \\
\lnl{line:co-exec4}   \tb {\bf case}  $V$ contains $m_2(v)$ from  $2t+1$ different processors: \hfill\textit{/* process the sender's message */}\\
\lnl{line:r1a}   \due {\bf if } $r>1$ {\bf wait until} $\;\forall q\in v, \tri{r-1,q}\in \bar\M$; \hfill\textit{/*   wait for the causally prior messages   */} \\
\nl \due{\bf Accept:} {\bf if}  $p\not=s $ {\bf then} $\M:=\M\cup\tri{r,s,v}$.\hfill\textit{/*   accept message $v$ from processor $s$ sent in round $r$  */} \\
%

%
%
\caption[caption]{ \cosend\!$(r,s)$: A casually ordered reliable broadcast \\\hspace{\textwidth}\mbox{\ \hspace{0.841in}} with asynchronous Byzantine faults for $n>3t$ }\label{figure:bb}
\end{algorithm}

Observe that the  simulation, presented in \algref{figure:state-sim}, produces per each processor an agreed upon sequence of sets of values  $M_0, ... M_{r-1}$ received by it's SM in the related rounds, thus, simulating the exact behavior of protocol $P$. This implies that the above simulation is a protocol to simulate in a $t$ resilient asynchronous Byzantine system a deterministic message passing protocol, $P$, in a \MAOBt system.

The delicate points in the simulation reside in the details of \cosend, which we now describe. 
The \cosend\ protocol, \algref{figure:bb}, is invoked per processor per sending round and consists of 5 conceptual steps.   
Initially (step 1) the sender of the current instance of the protocol sends its initial value to everyone.  
Thus, everyone should wait to receive the appropriate initial value. 
Due to asynchrony it may take time, but without faults, it would eventually arrive to everyone.  
Because of maliciousness, the message may not arrive to every processor.  Moreover, conflicting values might be sent to different processors.  The following steps intend to address exactly these difficulties.

If a processor receives an initial value (step 2) it notifies every processor by sending  $m_1(v)$ message.
Malicious behavior may cause different processors to send  $m_1$ messages for different values. 
Each processor sends at most a single  $m_1$ message per invocation of the protocol (per round).
A processor may receive several $m_1$ messages, even if it did not receive an initial value. 

In the 3rd step, a processor that has received $n-t$  identical copies of  $m_1$ messages for the same value, sends $m_2$ message. 
Notice that if the original sender is correct, this will eventually happen at every processor.
Observe, that no two correct processors send  $m_2$ messages with conflicting values since the protocol instructs a correct processor to send at most a single  $m_1$ message,  the $n-t$ threshold prevents two correct processors from getting $n-t$  copies of $m_1$ messages for different values.
Notice that a processor may receive several  $m_2$ messages without receiving $n-t$  copies of $m_1$ messages, if it receives at least $t+1$  $m_2$ messages it knows that at least one correct processor have sent one, so it can also join that by sending an  $m_2$ message (potentially skipping step 2 on the way).

To complete this part of the protocol a processor (step 4) waits to receive $2t+1$  $m_2$ messages. Once it receives that many identical $m_2$ messages it knows that eventually every processor will receive at least $t+1$, will send a  $m_2$ message to everyone else, which leads to everyone eventually receiving $2t+1$  $m_2$ messages. 

For $r=1$, the round of exchanging the input values, this completes the protocol. In all future rounds, there is an additional step (step 5) of waiting for all messages that are causally prior to the current message to be accepted (and processed), before the current message will be accepted. The content of the message, $v$, specifies explicitly the set of prior messages we need to wait for. If the sender is correct this will eventually happen at every processor. 
If the sender is faulty and claimed to receive messages from a processor that never sent it a message, the waiting for prior messages might not end, and the state machine of that (faulty) sender will practically  be blocked at every processor.  
However, if any correct will agree to accept the message, eventually everyone will receive all the prior messages and will accept the message.

Given the above discussion, it is clear that if the sender is correct, all processors eventually will complete the protocol and will accept its value.
Moreover, if any correct processor accepts a message, every correct will end up eventually accepting the same message, after accepting all causally prior messages to that message.


\begin{lemma}\label{lem:co}
For $n>3t,$
\algref{figure:bb} implements 
a Causally Ordered Reliable Broadcast transport layer in which a sender $s$ uses  \cosend to send its messages and  each processor accepts messages that satisfies:
\beginsmall{enumerate}
\item[\textup{ CO1:}]
If a correct sender, s, sends a consecutive sequence of messages, then every processor accepts the sequence in the same order that it was sent.

\item[\textup{ CO2:}] For $r>1$, if a processor, $p$, accepts  a $v$  via $\cosend(r,s)$, it already accepted $v_j$  via $\cosend(r-1,p_j)$, for every $p_j\in v.$

\item[\textup{ CO3:}]
If a processor, $p$, accepts  a $v_1$  via $\cosend(r,s)$ followed by a $v_2$  via $\cosend(r+1,s)$, then  any other processor $q$ will end up  accepting $v_1$ followed by $v_2$. 

\endsmall{enumerate}
\end{lemma}

The above discussions, \lemmaref{lem:co} and given that a processor moves to the next round, once it accepts and processes some $n-t$ current round messages implies the following result.

\begin{lemma}\label{lem:maobt}
Given  a deterministic protocol 
$P$ that is viewed as a function $\F(M_{r-1},S_p,r)$, for $r\ge 1$, 
in a \MAOBt system,
the protocol presented in \algref{figure:state-sim}  
simulates it  in a $t$ resilient asynchronous Byzantine system, given that $n>3t$.
\end{lemma}

\section{Simulating a protocol in a MOBtt system}

To finalize the main result of the paper we will now expand the simulation from simulating a 
protocol $P$ that runs is a  \MAOBt  system to a protocol in a   \MOBtt  system.  The extension is to ensure that before a processor adapts the $accept$ set it communicates with others to converge to $accept$ sets such that all sets have at least $n-t$ processors in common.
To achieve that we introduce a
third technique,  we run a couple of  rounds of the equivalent to a full information message exchange to make sure that everyone shares messages from a set of at least $n-t$ processors.  Once this happens the processor takes its next step.

The third primitive, \ccore, is an adaptation of the Get-Core approach mentioned in~\cite{DistComp} (attributed to the second Author) and a variation of it that was later presented in~\cite{Abraham2010} as Binding Gather, and using ideas from~\cite{Abraham2005}.

Each processor invokes the \ccore protocol, appearing in \algref{figure:ccore},  with a set of $n-t$  different processors IDs. Each correct processor, $p$,  returns as an output a set of at least $n-t$  different processors' IDs, such that at least $n-t$ of them are shared by the outputs of all correct processors. The \ccore  properties are:

\beginsmall{itemize}
\item {\em Validity}: At each correct processor, the output set of IDs contains the input set of IDs.
\item {\em Commonality}: There exists a set of $n-t$ IDs  that appears in the output set of every correct processor.
\item {\em Termination}: All correct processors eventually output some non-empty set of IDs.
\endsmall{itemize}

A set that is in every output set is called a common core. 
The \ccore primitive is  described in  \algref{figure:ccore}. In the first round, everyone sends its $accept[r]$ set.  
In the background the processor continues to update its $accept[r]$ set with messages it continues to accept.
To complete the first round of \ccore, it waits to receive at least $n-t$ sets that are contained in its current state of the set $accept[r]$.  This will eventually happen due to the \cosend properties, and the fact that there are at least $n-t$ correct processors.  Once this happens it sends again its current set and waits again to received at least $n-t$ second round sets that are contained in its current state of the set.

\begin{algorithm}[!ht]
\footnotesize
\SetNlSty{textbf}{}{.}
\setcounter{AlgoLine}{0}
 \ \hfill\textit{/* the input set $accept[r]$ is updated contineously in the background according to the messages accepted  via \cosend and processed in  \algref{figure:state-sim} */}\\
 \ \\
\lnl{line:cc-step1} \  {\bf step 1}  $send(r,1,accept[r])$ to all;\mbox{\ }\hfill\textit{/* send the input set to all, a processor sends also to itself */}\\
\lnl{line:cc-wait1} \  \tb  {\bf wait until} $|\{j\mid received(r,1,\pi_j) \mbox{ from } p_j, \mbox { and } \pi_j\subseteq accept[r] \}|\ge n-t;$\\
\ \\
\lnl{line:cc-step2} \  {\bf step 2}  $send(r,2,accept[r])$ to all;\mbox{\ }\hfill\textit{/* the set $accept[r]$ is being continuously updated in the background */}\\
\lnl{line:cc-wait2} \  \tb  {\bf wait until} $|\{j\mid received(r,2,\pi_j) \mbox{ from } p_j, \mbox { and } \pi_j\subseteq accept[r] \}|\ge n-t;$\\
\ \\
\lnl{line:cc-return}       {\bf return}  $accept[r]$. \\
\caption[caption]{$\ccore(accept[r])$, the Common Core protocol  }\label{figure:ccore}
\end{algorithm}

The correctness proof is similar to the proof of get-core in \cite{DistComp}. The original proof did not include Byzantine nodes, therefore we need to change it a bit.  Define a table $T$ with $n-t$ raws and $n-t$ columns, that refer to a set $\bar\Pi$ of $n-t$  correct  processors.  
The $accept[r]$ value of each correct processor contains at least $n-t$ IDs, therefore it contains at least $n-2t\ge t+1$ IDs of processors in $\bar\Pi$ represented in $T$.  
For $p_i,p_j\in \bar\Pi$, entry $T[i,j]$ in the table is $1$ if $p_j$ is one of the $n-t$ processors that $p_i$ waited for in order to complete the first round of \ccore, and $0$ otherwise.  Observe that if $1$ appears in  entry $T[i,j]$, the $accept_i[r]$ sent by $p_i$ in the second round contains all the $n-t$ IDs appearing in 
the initial $accept_j[r]$ sent by $p_j$  in the first round of the \ccore protocol.

Since all processors in $\bar\Pi$ will eventually invoke \ccore, $T$ will contain  at least $(n-t)(t+1)$ entries with $1$. This implies that there is an ID of a correct processor, say $\bar p$, that appears in at least $t+1$ raws.  
Thus, there are at least $t+1$ correct processors whose second round set includes the $n-t$ IDs that appear in the initial set of $\bar p$.  
Before completing the protocol, each processor waits to get the sets of $n-t$ processors, so it includes the set of at least one of these $t+1$ processors, thus includes the set of $n-t$  IDs appearing in the initial list of $\bar p$.

\begin{lemma}\label{lem:ccore}
For $n>3t$, the protocol presented in \algref{figure:ccore}  
implements the \ccore properties. 
\end{lemma}

To obtain the  final protocol we add the \ccore invocation to the simulation protocol presented in \algref{figure:state-sim}.
We invoke the \ccore protocol on all the  $accept$ sets of a given round after completing \lref{line:collect} and before executing \lref{line:send} of \algref{figure:state-sim}. The output of the \ccore is used in \lref{line:send}  as the set of processors from which we received messages from in that round. \algref{figure:full-state-sim} presents the complete protocol.

\begin{algorithm}[!ht]
\footnotesize
\SetNlSty{textbf}{}{.}
\setcounter{AlgoLine}{0}
\lnl{line:f-def}  \     {\bf set}  $\forall k$ $accept[k]:=\emptyset$; \hspace{2mm} \hfill\textit{/*  the sets of accepted senders at various rounds; executed at processor $p$ */}\\
\lnl{line:f-def2}  \     {\bf set}  $\bar\M:=\emptyset$;\hspace{2mm} \hfill\textit{/*  the set of processed messages */}\\
\lnl{line:f-def3}  \     {\bf set}  $\M:=\emptyset;$\hspace{2mm} \hfill\textit{/*  the set of accepted  messages that were not processed yet  */}\\
\ \\
\lnl{line:f-step1} \     {\bf invoke }  $\cosend(1,p)$ to broadcast $\I$;\mbox{\ }\hfill\textit{/* broadcast the input value, a processor sends also to itself */}\\
\lnl{line:f-init-r} \    $r:=1;$\hfill\textit{/* the round number */}\\
\ \\
\nl \    {\bf do }  {\bf until} $SM_p$ halts:\\
\lnl{line:f-collect}\    \tb {\bf wait until } $|accept[r]|\ge n-t$ and $p\in accept[r]$; \hfill\textit{/* participate in all $\cosend(\ell,*),$ $\ell \le r$, protocols */}\\
\lnl{line:f-p-core}\  \tb  $accept[r]:= \ccore(accept[r])$ \hfill\textit{/* the 2 rounds protocol to converge to shared $n-t$ */}\\
\lnl{line:f-send} \    \tb {\bf invoke } $\cosend(r+1,p)$ to broadcast $accept[r]$;\mbox{\ }\hfill\textit{/* broadcast the accepted set in round $r$ */}\\%
\lnl{line:f-inc}\    \tb $r:=r+1$;\\
\lnl{line:f-end} \    {\bf end.} \\
\ \\
\lnl{line:f-r1}   {\underline {\bf In the Background:}} Execute  for each $\tri{r',p_i,\pi}\in\M$: \hfill\textit{/*   message received via \cosend from $p_i$ for round $r$   */} \\
\lnl{line:f-init-rec}    \tb {\bf if} $r'=1$ {\bf then start } $SM_i$ with input $\pi$;\hfill\textit{/* start a SM with the initial input */}\\
\lnl{line:f-init-rec2}     \tb {\bf if} $r'>1$ {\bf then } \\
\lnl{line:f-rec-m}  \tb\tb {\bf let} $M:=\{m_j\mid p_j\in\pi \mbox{ and } SM_j[r'-1] \mbox{ sends } m_j \mbox{ to } p_i\}$; \hfill\textit{/* the messages $p_i$ should have received */}\\
\lnl{line:f-apply-m}  \tb\tb$SM_i[k]:=\F(M,SM_i[r'-1],r')$; \hfill\textit{/* apply  protocol $\F$ to determine the next state of $SM_i$ */}\\
\lnl{line:f-r3}    \tb $\M:=\M\setminus \tri{r',p_i,\pi}$;\\
\lnl{line:f-r4}     \tb $\bar\M:=\bar\M\cup \tri{r',p_i}$;\\
\lnl{line:f-add}       \tb   $accept[r']:=accept[r']\cup\{p_i\}$. \\
\caption[caption]{Simulating a deterministic protocol of a \MOBtt system  \\\hspace{\textwidth}\mbox{\ \hspace{0.83in}} in an asynchronous Byzantine system with $n>3t$}\label{figure:full-state-sim}
\end{algorithm}

\begin{theorem}\label{thm:full}
Given  a deterministic protocol 
$P$ that is viewed as a function $\F(M_{r-1},S_p,r)$, for $r\ge 1$, 
in a \MOBtt system,
the protocol presented in \algref{figure:full-state-sim}  
simulates it  in a $t$ resilient asynchronous Byzantine system, given that $n>3t$.
\end{theorem}

\begin{corollary}\label{col:full}
For $n>3t$ and deterministic protocols, 
the $t$ resilient asynchronous Byzantine system and 
the \MOBtt system are equivalent.
\end{corollary}



\newpage

\bibliographystyle{plain}
\bibliography{bibliography}

\begin{thebibliography}{10}

\bibitem{Abraham2010}
Ittai Abraham, Marcos~K. Aguilera, and Dahlia Malkhi.
\newblock Fast asynchronous consensus with optimal resilience.
\newblock In Nancy~A. Lynch and Alexander~A. Shvartsman, editors, {\em
  Distributed Computing: 24th International Symposium, DISC 2010, Cambridge,
  MA, USA, September 13-15, 2010. Proceedings}, pages 4--19, Berlin,
  Heidelberg, 2010. Springer Berlin Heidelberg.

\bibitem{Abraham2005}
Ittai Abraham, Yonatan Amit, and Danny Dolev.
\newblock Optimal resilience asynchronous approximate agreement.
\newblock In Teruo Higashino, editor, {\em Principles of Distributed Systems:
  8th International Conference, OPODIS 2004, Grenoble, France, December 15-17,
  2004, Revised Selected Papers}, pages 229--239, Berlin, Heidelberg, 2005.
  Springer Berlin Heidelberg.

\bibitem{DiskPaxos2005}
Ittai Abraham, Gregory Chockler, Idit Keidar, and Dahlia Malkhi.
\newblock Byzantine disk paxos: optimal resilience with byzantine shared
  memory.
\newblock {\em Distributed Computing}, 18(5):387--408, 2005.

\bibitem{AG15}
Yehuda Afek and Eli Gafni.
\newblock A simple characterization of asynchronous computations.
\newblock {\em Theor. Comput. Sci.}, 561:88--95, 2015.

\bibitem{Aiyer2007}
Amitanand~S. Aiyer, Lorenzo Alvisi, and Rida~A. Bazzi.
\newblock Bounded wait-free implementation of optimally resilient byzantine
  storage without (unproven) cryptographic assumptions.
\newblock In Andrzej Pelc, editor, {\em Distributed Computing: 21st
  International Symposium, DISC 2007, Lemesos, Cyprus, September 24-26, 2007.
  Proceedings}, pages 7--19, Berlin, Heidelberg, 2007. Springer Berlin
  Heidelberg.

\bibitem{DistComp}
H.~Attiya and J.~Welch.
\newblock {\em Distributed Computing: Fundamentals, Simulations and Advanced
  Topics}.
\newblock John Wiley \& Sons, 2004.

\bibitem{BH2007}
Zuzana Beerliov{\'a}-Trub{\'i}niov{\'a} and Martin Hirt.
\newblock {\em Simple and Efficient Perfectly-Secure Asynchronous MPC}, pages
  376--392.
\newblock Springer Berlin Heidelberg, Berlin, Heidelberg, 2007.

\bibitem{BenOr83}
Michael Ben-Or.
\newblock Another advantage of free choice (extended abstract): Completely
  asynchronous agreement protocols.
\newblock In {\em PODC '83: Proceedings of the second annual ACM symposium on
  Principles of distributed computing}, pages 27--30, New York, NY, USA, 1983.

\bibitem{Ben-Or:1993}
Michael Ben-Or, Ran Canetti, and Oded Goldreich.
\newblock Asynchronous secure computation.
\newblock In {\em Proceedings of the Twenty-fifth Annual ACM Symposium on
  Theory of Computing}, STOC '93, pages 52--61, New York, NY, USA, 1993. ACM.

\bibitem{Bracha:1984}
Gabriel Bracha.
\newblock An asynchronous [(n - 1)/3]-resilient consensus protocol.
\newblock In {\em Proceedings of the Third Annual ACM Symposium on Principles
  of Distributed Computing}, PODC '84, pages 154--162, New York, NY, USA, 1984.
  ACM.

\bibitem{Bracha:1987}
Gabriel Bracha.
\newblock Asynchronous byzantine agreement protocols.
\newblock {\em Information and Computation}, 75(2):130 -- 143, 1987.

\bibitem{coan88}
B.~A. Coan.
\newblock A compiler that increases the fault tolerance of asynchronous
  protocols.
\newblock {\em IEEE Transactions on Computers}, 37(12):1541--1553, Dec 1988.

\bibitem{hybrid-arxiv}
Danny Dolev and Eli gafni.
\newblock Synchronous hybrid message-adversary.
\newblock Technical report, ArXiv, May 2016.
\newblock url: http://arxiv.org/abs/1605.02279.

\bibitem{Rachid}
Eli Gafni and Rachid Guerraoui.
\newblock Generalized universality.
\newblock In Joost{-}Pieter Katoen and Barbara K{\"{o}}nig, editors, {\em
  {CONCUR} 2011 - Concurrency Theory - 22nd International Conference, {CONCUR}
  2011, Aachen, Germany, September 6-9, 2011. Proceedings}, volume 6901 of {\em
  Lecture Notes in Computer Science}, pages 17--27. Springer, 2011.

\bibitem{Malkhi98DC}
Dahlia Malkhi and Michael Reiter.
\newblock Byzantine quorum systems.
\newblock {\em Distributed Computing}, 11(4):203--213, 1998.

\bibitem{Mendes2015}
Hammurabi Mendes, Maurice Herlihy, Nitin Vaidya, and Vijay~K. Garg.
\newblock Multidimensional agreement in byzantine systems.
\newblock {\em Distributed Computing}, 28(6):423--441, 2015.

\bibitem{Mendes2014}
Hammurabi Mendes, Christine Tasson, and Maurice Herlihy.
\newblock Distributed computability in byzantine asynchronous systems.
\newblock In {\em Proceedings of the 46th Annual ACM Symposium on Theory of
  Computing}, STOC '14, pages 704--713, New York, NY, USA, 2014. ACM.

\bibitem{neiger88}
Gil Neiger and Sam Toueg.
\newblock Automatically increasing the fault-tolerance of distributed systems.
\newblock In {\em PODC '88: Proceedings of the seventh annual ACM Symposium on
  Principles of distributed computing}, pages 248--262, New York, NY, USA,
  1988.

\bibitem{FTState90}
F.~B. Schneider.
\newblock Implementing fault-tolerant services using the state machine
  approach: a tutorial.
\newblock {\em ACM Computing Surveys (CSUR)}, 22(4):299--319, 1990.

\bibitem{Srikanth87}
T.~K. Srikanth and Sam Toueg.
\newblock Simulating authenticated broadcasts to derive simple fault-tolerant
  algorithms.
\newblock {\em Distributed Computing}, 2(2):80--94, 1987.

\bibitem{TV14}
Lewis Tseng and Nitin~H. Vaidya.
\newblock Asynchronous convex hull consensus in the presence of crash faults.
\newblock In {\em Proceedings of the 2014 ACM Symposium on Principles of
  Distributed Computing}, PODC '14, pages 396--405, New York, NY, USA, 2014.
  ACM.

\end{thebibliography}

\appendix

\noindent {\Large \bf Appendix}\label{sec:appendix}

\section{Simulating Probabilistic Protocols}
In order to extend the DG-simulation above to randomized protocols one can't just ask a processor to distribute together with its collected set its random choice for the current round. The reason is that Byzantine processors may not draw the random bits from the expected distribution.  To deal with that all processors collectively  choose the random choices for each processor once it broadcasts its candidates' set.

For simplicity, we assume that the protocol instructs each processor to flip a fair coin at the beginning of each round. The same construction can be extended to every other required distribution of random values.  What we show now is a method to produce for each simulated processor a 
coin flip with an arbitrary small bias.

In case $n>4t$ one can compute any probabilistic function using Asynchronous Multi-Party-Computation,  in the presence of up to $t$ Byzantine faults, see~\cite{Ben-Or:1993,BH2007}.  
We make use of such a protocol. The protocol $\simcoin(s)$ (\algref{figure:c-state-sim2}) returns a shared coin $c$ on behalf of a pre-specific processor $p_i$ at all participating processors. The main difference from previous protocols is \lref{line:c-coin2} that obtains the collective coin, and in the next line we apply the result to the state machine of processor $p_i$.

The question of whether there is a small bias coin flipping protocol for $n>3t$ that runs is a constant number of rounds is an open question.

\begin{algorithm}[!ht]
\footnotesize
\SetNlSty{textbf}{}{.}
\setcounter{AlgoLine}{0}
\lnl{line:c-def}  \     {\bf set}  $\forall k$ $accept[k]:=\emptyset$; \hspace{2mm} \hfill\textit{/*  the sets of accepted senders at various rounds; executed at processor $p$ */}\\
\lnl{line:c-def2}  \     {\bf set}  $\bar\M:=\emptyset$;\hspace{2mm} \hfill\textit{/*  the set of processed messages */}\\
\lnl{line:c-def3}  \     {\bf set}  $\M:=\emptyset;$\hspace{2mm} \hfill\textit{/*  the set of accepted  messages that were not processed yet  */}\\
\ \\
\lnl{line:c-step1} \     {\bf invoke }  $\cosend(1,p)$ to broadcast $\I$;\mbox{\ }\hfill\textit{/* broadcast the input value, a processor sends also to itself */}\\
\lnl{line:c-init-r} \    $r:=1;$\hfill\textit{/* the round number */}\\
\ \\
\nl \    {\bf do }  {\bf until} $SM_p$ halts:\\
\lnl{line:c-collect}\    \tb {\bf wait until } $|accept[r]|\ge n-t$ and $p\in accept[r]$; \hfill\textit{/* participate in all $\cosend(\ell,*),$ $\ell \le r$, protocols */}\\
\lnl{line:c-p-core}\  \tb  $accept[r]:= \ccore(accept[r])$ \hfill\textit{/* the 2 rounds protocol to converge to shared $n-t$ */}\\
\lnl{line:c-send} \    \tb {\bf invoke } $\cosend(r+1,p)$ to broadcast $accept[r]$;\mbox{\ }\hfill\textit{/* broadcast the accepted set in round $r$ */}\\%
\lnl{line:c-inc}\    \tb $r:=r+1$;\\
\lnl{line:c-end} \    {\bf end.} \\
\ \\
\lnl{line:c-r1}   {\underline {\bf In the Background:}} Execute  for each $\tri{r',p_i,\pi}\in\M$: \hfill\textit{/*   message received via \cosend from $p_i$ for round $r$   */} \\
\lnl{line:c-init-rec}    \tb {\bf if} $r'=1$ {\bf then start } $SM_i$ with input $\pi$;\hfill\textit{/* start a SM with the initial input */}\\
\lnl{line:c-init-rec2}     \tb {\bf if} $r'>1$ {\bf then } \\
\lnl{line:c-rec-m}  \tb\tb {\bf let} $M:=\{m_j\mid p_j\in\pi \mbox{ and } SM_j[r'-1] \mbox{ sends } m_j \mbox{ to } p_i\}$; \hfill\textit{/* the messages $p_i$ should have received */}\\
\lnl{line:c-coin2} \tb\tb {\bf wait until:} $\tri{c}:=\simcoin(p_i)$; \hfill\textit{/* obtain a coin for $p_i$ */}\\
\lnl{line:c-apply-m} \tb\tb$SM_i[k]:=\F(M,c,SM_i[r'-1],r')$; \hfill\textit{/* apply  protocol $\F$ to determine the next state of $SM_i$ */}\\

\lnl{line:c-r3}    \tb $\M:=\M\setminus \tri{r',p_i,\pi}$;\\
\lnl{line:c-r4}     \tb $\bar\M:=\bar\M\cup \tri{r',p_i}$;\\
\lnl{line:c-add}       \tb   $accept[r']:=accept[r']\cup\{p_i\}$. \\
\caption[caption]{Simulating a deterministic protocol of a \MOBtt system  \\\hspace{\textwidth}\mbox{\ \hspace{0.83in}} in an asynchronous Byzantine system with $n>3t$}\label{figure:c-state-sim2}
\end{algorithm}

\end{document}